\documentclass[preprint,showpacs,preprintnumbers,amsmath,amssymb]{revtex4}
\usepackage{graphicx}% Include figure files
\usepackage{dcolumn}% Align table columns on decimal point
\usepackage{bm}% bold math

\begin{document}

\preprint{}

\title{ Effects of elasticity on the shape of measured shear signal in two-dimensional assembly of disks}
\author{O. Pozo$^1$ and N.Olivi-Tran$^2$}
\affiliation{$^1$LPMC, UMR-CNRS 6622, Parc Valrose,Universite de Nice Sophia Antipolis,  06108 Nice cedex 2, France, \\ $^2$SPCTS, UMR-CNRS 6638, Ecole Nationale Superieure de
        Ceramiques Industrielles,47 av. A.Thomas, 87065 Limoges cedex, France}

\date{\today}% No date.

\begin{abstract}
A Molecular Dynamics approach has been used to compute the shear force
resulting from the shearing of disks.
Two-dimensional monodisperse disks have been put in an horizontal and rectangular shearing cell with
periodic boundary conditions on right and left hand sides.
The shear is applied by pulling the cover of the cell either  at a constant
rate or by pulling a spring, linked to the cover, with a constant force.
Depending on the rate of shearing and on the elasticity of the whole set up, we showed that
the  measured shear force signal is either irregular in time, regular in time
but not in shape, or regular in shape.
\end{abstract}
%Use showkeys class option if keyword
                              %display desired

%%%%%%%%%%%%%%%%%%%%%%%%%%%%%%%%%%%%%%%%%%%%%%%%%%%%%%%%%%%%%%%%%%%

%  To kick in the changes.
% Make title here for two column format.
\maketitle
\frenchspacing   % no double spaces after colon
                 % added by <W.Hennings@fz-juelich.de>

% Write some ascii text files called intro.tex, concept.tex, etc.
% TeX and LaTeX will look for the .tex subscript by default.
\section{Introduction}
When a macroscopically flat solid is pulled horizontally over an also macroscopically
flat substrate, for given rates of pulling and of normal loads, the solid undergoes an intermittent behavior made of periods
where it sticks to the substrate followed by periods where it slips over
the same substrate. Several models have been written to explain
this phenomenon.  
The first model explained this stick slip behavior by  parameters
of the system: the static friction and the dynamic friction coefficients
and the normal load
\cite{coulomb}.
In the years 1950, Bowden and Tabor made the theory that the friction
coefficient depends on the strength of the material \cite{bowden}.
Moreover, Dieterich \cite{dieterich1,dieterich2} and Heslot et al. \cite{heslot}
demonstrated that the friction coefficients where not constant during
one experiment, they showed that there was an aging process 
in the material.
Finally, Baumberger \cite{baumberger} showed that the transition from a continuous sliding to a stick slip behavior was dependent on the elasticity
of the measuring set up.
A model was introduced by Dieterich, Rice and Ruina \cite{dieterich,ruina}
to give a mathematical basis to the stick slip phenomenon.

By the same way, if a granular medium  is sheared
by pulling an horizontal cover on the top of it, the granular system
undergoes alternatively sticks (i.e. the whole system is stopped)
and slips. This stick slip behavior depends on the rate at which
the granular medium is pulled and on the normal load which is applied
on it.
Several experiments have been made on granular matter under shear
as well for dry grains \cite{nasuno,nasuno2,lubert,deryck,coste,albert,geng} as for cohesive grains \cite{oliver}.Numerical simulations have been performed on this topic by Thompson and Grest
\cite{thompson}.

But the question of the elasticity of all the elements setting up the shear
experiment has almost not been studied. Baumberger \cite{baumberger} has shown that
the existence of a stick slip behavior when dragging a solid inside a granular medium,
depends on the elasticity of the elastic component of the pulling set up.
However, it is very difficult to estimate the elasticity of the shearing
apparatus, and therefore  of the grains and in case of cohesive grains, the elasticity
of the interactions between neighbouring grains.
A previous study \cite{olivi} has shown that in the case of a numerical model
of the shearing, at a constant shearing rate, of two-dimensional disks ,
it is possible to replace the action of the grains by one single
spring. But what happens if all the components of the experimental or numerical
set up have their own elasticity?
In order to answer this question, we computed by a numerical model
the shearing of disks in a shearing cell with non infinite and infinite
elasticity, and with cohesion or without cohesion between neighbouring
grains.

\section{Numerical Model}

Molecular Dynamics is a powerful numerical method                               to study the dynamics of granular materials \cite{truc}.
The model we used here is a  version of Molecular                               Dynamics for granular flow with cohesion in a two-dimensional
shearing cell \cite{olivi,truc,truc2,olivi1,olivi2}.           
  Particles are modeled as N disks that have equal density $\rho=2.2g.cm^{-2}$ and
 diameters $d=200\mu m$.
        The only external force acting on the system results from               gravity, $g = 981 cm.s^{-2}$.
        The particle-particle and particle-wall contacts are                    described in the normal direction (i.e. in the particles'
center-center direction) by a Hooke-like force law. The normal force            is written:
\begin{eqnarray}
{\bf f}_n(i,j)=(-Yr_{eff}[\frac{1}{2} (d_i+d_j)-|{\bf r}_{i,j}|]+\nonumber \\ \gamma \frac{m_{eff}({\bf v}_{i,j}.{\bf r} _{i,j})}{r_{eff}|{\bf r}_{i,j}|})\frac{{\bf r}_{i,j}}{| {\bf r}_{i,j}|}
\end{eqnarray}
where $Y$ is the Young modulus of the solid, $r_{eff}$ ($m_{eff}$) stands for the
effective radius (mass) of the particles $i$ and $j$, ${\bf v_{ij}}={\bf v_i}- {\bf v_j}$ is the relative
velocity of particle $j$ towards particle $i$.
The effective radius is defined here as the radius of the disk which effectively
enters in the interaction with another disk, it can be smaller than the real
disk radius in the case where the disks overlap or larger in the case of
capillary interaction (see below). The effective radius is proportional to the distance
between the centers of the disks if these two are in interaction: 
\begin{equation}
r_{eff}=\frac{(|({\bf r_i}-{\bf r_j})|)^2}{(d_i+d_j)/4}
\end{equation}
assuming that the disks never overlap more than $d_i/4$ and therefore adapting the time step.

The effective mass is defined here
as the mass which contributes to the shearing, in our case all disks have their effective mass equal to their real mass except the disks which build the bottom
of the shearing cell.
$d_i$ (resp. $d_j$) is the diameter of particle $i$ (resp. $j$) and ${\bf r_{ij}}$
points from particle $i$ to particle $j$.
$\gamma$ is a phenomenological dissipation coefficient.

        We model the  friction force between particles by
putting a virtual spring at the point of first contact. Its
elongation is integrated over the entire collision time and set to
zero when the contact is lost. The maximum possible value of the
restoring force in the shear direction (i.e. in the plane
perpendicular to the normal direction), according to Coulomb's
criterion, is proportional to the normal force multiplied by the
friction coefficient $\mu$. It gives a friction force ${\bf f_s}(i,j)$ which is
written:
\begin{eqnarray}
{\bf f}_s (i,j)=-sign(f_f(i,j))min(f_f(i,j),\mu|f_n(i,j)|){\bf s}\\
{\bf f}_f(i,j):= - \int (\dot{r}_i- \dot{r}_j) {\bf s} dt
\end{eqnarray}
where $s$ stands for the unit vector in the shear direction.
        When a particle collides with the cylinder wall, the same
forces (1) and (2) act with infinite mass and radius for particle $j$.
The disks are allowed to rotate. The rotational force is given
by:
\begin{equation}
{\bf f}_{rot}=r_{eff}{\bf f}_s
\end{equation}

 Cohesive forces were modelled by adding a spring
force to the normal force when particles are in contact:
\begin{equation}
f_{cap}=Kr_{eff}\frac{1}{2} (d_i+d_j)
\end{equation}
where $K$ is the corresponding elastic constant (with a dimension
of a Young modulus).
When the distance (proportional to the effective radius) between the surface of the particles is lower than 10 \% of the diameter
of the smallest particle, the value of the spring constant is multiplied by the distance
between the  particles. This additive
force is set to zero when the elongation of the virtual spring
reaches a maximum length of 10\% of the smallest particle diameter.

This cohesive force is a good model for capillary forces
between beads with nanoasperities.

The disks are put in an horizontal and  rectangular shearing cell with or without blades, one can see an example
of this shearing cell in ref.\cite{olivi}. Periodic boundary conditions
are imposed on the left and right hand side of the cell.
Shear is applied on the disks in the shearing cell by translating
the upper cover on which are linked blades to drag the disks or without any blades.
The shear cell is made of disks in order to simplify
the interactions of the bulk disks and the shear cell: the upper cover is made of disks
of non zero effective mass (allowing the vertical move of the cover)
and the bottom of the cell is made of disks with zero effective mass
(allowing a stable position of the shear cell with respect to gravity).
Shear is applied on the disks by translating the upper cover either at
constant velocity, whatever the resistance to translation
of the granular medium, or with a spring linked to the cover. In this last
case, the acceleration ${\bf a_c}$ and force ${\bf f_c}$ acted by the cover on the disks assembly
is given by:
\begin{eqnarray}
{\bf a_c}(t+dt)=({\bf f_c}(t)+K_c{\bf x}(t))/p_w \\
{\bf x}(t+dt)={\bf x}(t)+K_c{\bf x}(t)
\end{eqnarray}
where $K_c$ is the elasticity constant of the spring pulling the cover and $p_w$ is the weight of the cover.

The upper cover of the cell is allowed to undergo a vertical shift,
the magnitude of this shift depending on the weight of this upper cover ( and therefore
on the effective mass of the upper cover disks), on the velocity of the upper cover and on the disks assembly dilation.
We added no vertical elasticity to the cover as it would be more difficult
to analyze the results with one more elastic component.

\section{Results and Discussion}
The parameters of our computations
were the followings:   the weight of the cover was equal to $=0.552g$, the mass of one disk
being equal to $2.76.10^{-3}g$. The length of the cell was $L=8mm$ and the height
of the blades
was $h=1mm$. We used $N=720$ disks.
The dissipation coefficient was equal to $0.7m.s^{-1}$, the Young modulus was
$Y=14640g.cm^{-1}.s^{-2}$ and the spring constant was $K=300g.cm^{-1}.s^{-2}$
for the case of cohesive disks.
We computed the total force acted on the upper cover by the disk
assembly as a resistance to shear. For this, we added only the
horizontal coordinates of the forces acting on the cover.
This total force is similar to a stress.
The height of the cover  increased after the beginning of the simulation
for all cases of shear rates and elastic constants.

Results are given in figures 1, 2 , 3 and 4. These figures represent different
responses of the numerical set up as a function of the cover elasticity
(vertical coordinates) and shearing rates (horizontal coordinates).
Each symbol corresponds to one value of the cover elasticity and one
value of the shearing rate, each of them given by its coordinates. For the sake of simplicity, we gave an elasticity
of 100$g.s^{-2}$ to covers with infinite elasticity as our plots are in log-log.
Five regions have been identified in these plots:
\begin{itemize}
\item I: regular in time (with a characteristic frequency) but irregular in shape (noisy)
\item II: irregular in time and irregular in shape
\item III: curved slip signal and irregular in time
\item IV: regular in shape and irregular in time
\item V: regular in shape and in time
\end{itemize}

Figure 1 presents the different behaviors of the shear signal for a shear cell without blades and for disks without cohesion as a function of the shear rate and of the elasticity constant of the cover.
Figure 2 presents the same feature for disks with cohesion.
Figures 3 and 4 correspond to a shear cell with blades for respectively
non cohesive and cohesive disks.
An example of different stick slip signals is presented in figure 5.
The black bold curve corresponds to region II, the long dashed curve corresponds
to region IV and the thin continuous curve corresponds to region III.

Let us first analyze the elasticity of the setup and disks leading to the behavior corresponding to region I in fig. 1,2,3 and 4.

Region I is uniquely present when the cover is pulled with an infinite
elasticity, i.e. when it slides at a constant rate.
The presence of cohesion between the disks does not change this behavior
neither the presence of blades on the cover.
This characteristic behavior is the one which has been studied in ref. \cite{olivi2}. In this previous study, the authors showed that the whole assembly
of disks could be replaced by one single spring in the case of cohesive
disks. Here, we have the same behavior (presence of a characteristic
frequency of the shear signal but irregular signal) even when there
is no cohesion between the disks. Straightforwardly, we can say that,
as the cohesion has the dimension of a Young modulus, it only modifies
the resulting (effective) Young modulus of the disks. Therefore,
the assembly of disks behaves as a single spring which is the only
source of elasticity in the experiment.

Now, considering region II , let us compare fig. 1,2,3 and 4.
In fig. 1, corresponding to a setup with a varying elasticity
and disks without cohesion, the shear signal is irregular in time
and in shape for all the values of the elasticity
of the cover but only for small shear rates. In fig. 2, this behavior disappears for shear rates
larger than 0.4$mm.s^{-1}$ for the same values of the elasticity of the cover.
In fig. 3 and 4, only elasticities of the cover larger than 10 lead to this
behavior, in the case  of shear rates smaller than 20$mm.s^{-1}$.
We can say that the three (resp. two) origins of elasticities for fig. 2
and 4 (resp. for fig. 1 and 3) lead to an incoherent shear signal
which in these cases has no characteristic frequency and is irregular
in time. The system in this case is equivalent to 2 or 3 springs in series, with
equivalent spring constant. The resulting shear signal is therefore incoherent.

Region III corresponds to shear signals which are smooth but irregular in time.
Moreover,  the shear signal corresponding to the stick and to the slip
stages are curved, showing a non linearity in the elasticity of the experiment.
Region III is absent in fig.1 and appears in fig. 2 3 and 4 for small
spring constant (equal to 1) of the cover.
Here again we can say that the experiment is equivalent to two
or three springs in series; but, contrarily to the behavior of the shear signal
of region II, the spring constant of the cover is small and becoming
dominant in the  behavior of the system (we recall that
the resulting spring constant of springs in series is the inverse
of the sum of inverses of each spring constant).
 
Region IV corresponds to shear signals which are regular in shape but
not regular in time (with no characteristic frequency).
As for  region III, the system in this case may be modelled by
two or three springs in series, the spring with the lowest spring constant
is dominant giving the behavior of the resulting shear signal.
The difference between
the spring constant of the cover and the one or two other spring constants
(representing the disks assembly and the cohesion) is smaller than
in the case of region III. 

In the case of region V, we can always consider that the system
is equivalent to two or three springs in series. We can consider that
there is a characteristic frequency of the shear signal
because the two or three spring constants have each a characteristic
frequency close from each other.

As a summary, let us calculate the spring constants of each of the components
of the system.
The spring constant $K_c$ of the cover is an input parameter of the computation.
The spring constant $K_d$ of the beads without cohesion (dry beads) may be given by:
\begin{equation}
K_d=\frac{1}{(\sum _{i=1,n} 1 / \frac{1}{\sum_{j=1,m} 1/K_{b1}})}=\frac{K_{b1}}{mn}
\end{equation}
where $K_{b1}$ is the spring constant directly related to the Young modulus $Y$
of one bead. $n$ corresponds to the mean number of disks per vertical line in the
shear cell while $m=40$ is the mean horizontal number of disks per horizontal line. This is equivalent to say that we have $n$ equivalent springs in series
each of these equivalent springs being one line of $m$ disks in series.

Similarly, the equivalent spring constant $K_h$ of cohesive disks (humid disks)
may be computed using the same formula but with a value of the equivalent spring constant $K_{b2}$
of each disks modified in order to take account of the cohesion and  of the Young modulus which can be considered as springs acting in parallel
$K_{b2}= (Y+K)d_i$. The equivalent spring constant of the assembly of disks
is then approximatively equal to $K_h=\frac{K{b2}}{mn}$.

Depending on the geometry of the shear cell, the values  of $n$
may vary. Indeed, in a shear cell with blades, only the disks which
are  below the blades enter the shearing process.
For the shear cell without blades, $n=18$ and for the shear cell with
blades $n=13$ corresponding to the the number of active disks
in the shear process, per vertical line.
Therefore, the actual values are for the cover $K_c=1,10, \infty g.s^{-2}$,for the shear cell
without blades  and for disks without cohesion $K_d=0.4 g.s^{-2}$ and for disks with cohesion $K_h=0.415g.s^{-2}$.
In the case of the shear cell with blades $K_d=0.56 g.s^{-2}$ and for disks with cohesion $K_h=0.57g.s^{-2}$.
The shear signal is regular in shape when $K_c$ is of the same order 
than $K_d$ or $K_h$ and for large shear rates, like in regions IV and V.
Otherwise, the shear signal is irregular in shape.

In ref. \cite{olivi}, we saw that the position of the spring corresponding
to the assembly of disks could be modelled by the following equation:
\begin{equation}
x(t)=\frac{v}{\omega}\sin(\omega t)
\end{equation}
where $v$ is the shear rate and $\omega=\sqrt{\frac{K_e}{M}}$, with $K_e$ the equivalent spring constant
of the disks and $M$ the mass of the disks assembly.
Here again we may use this equation for the whole system
assuming that the equivalent spring constant is $K_e=1/(\frac{1}{K_c}+\frac{1}{K_{h or d}})$ and $M$ is the sum of the mass of the cover and of the disks
 assembly. Therefore, one may see that if the value of  $v$
is much larger than $K_e$ the shear signal presents a characteristic 
frequency like in regions I and V.
If the value of $v$ is smaller than $K_e$, no characteristic frequency
appears like in regions II,III and IV.

Finally, we can say that the shape of the signal and its frequency
may be analyzed separately. The shape of the signal depends 
on the relative values of all components of the system
while the frequency of the shear signal depends on the mass, the shear
rate and the elasticity of the system.

\section{Conclusion}
We made a numerical simulations by molecular dynamics of the shearing
of twodimensional disks.
Depending on the elastic constant of each component of the set up
(of the cover, of the disks, and of the cohesion between the disks) and
of the shear rate,
the resulting simulated shear signal is either regular or irregular in shape,
and with or without  a characteristic frequency.
An analytic analysis allowed us to understand this behavior
in the light of the relative values of the elastic constants entering
the setup. It appears that we found a regular shear signal only
for given values of the elastic constants and of the shear rate.

\pagebreak
\begin{figure}
\includegraphics[width=14cm]{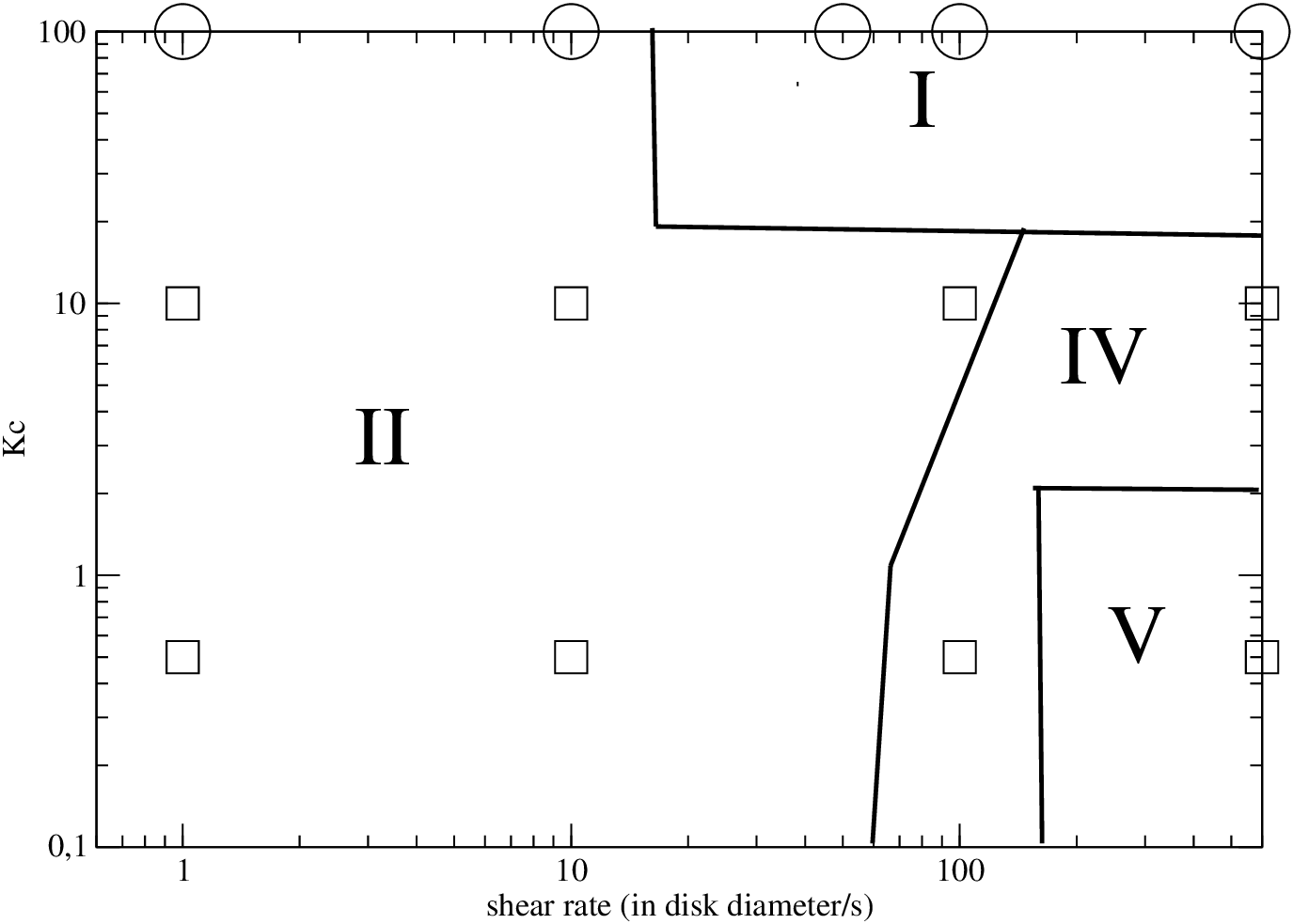}
\caption{Diagram of the computed shear signal for a shear cell withoud blades and for disks without cohesion
as a function of the elastic constant $K_c$ ($g.s^{-2}$) of the cover and of the shear rate (in disk diameter per $s^{-1}$)}
\end{figure}
\begin{figure}                                                                 
 \includegraphics[width=14cm]{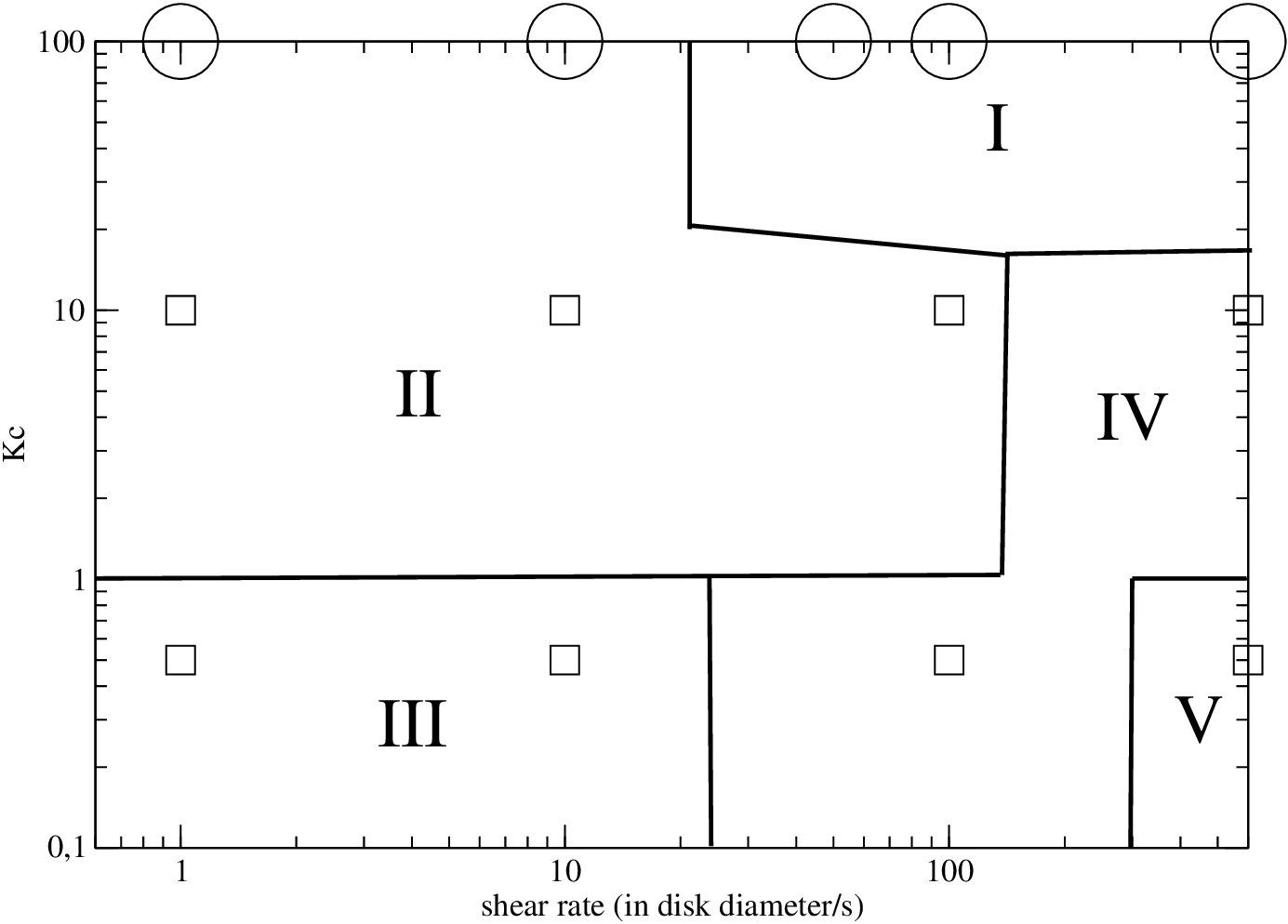}                                                      \caption{Diagram of the computed shear signal for a shear cell without blades
and for cohesive disks      as a function of the elastic constant $K_c$ 
( $g.s^{-2}$) of the cover and of the shear rate               (in disk diameter per $s^{-1}$)                                     }                                                                               \end{figure}

\begin{figure}   
  \includegraphics[width=14cm]{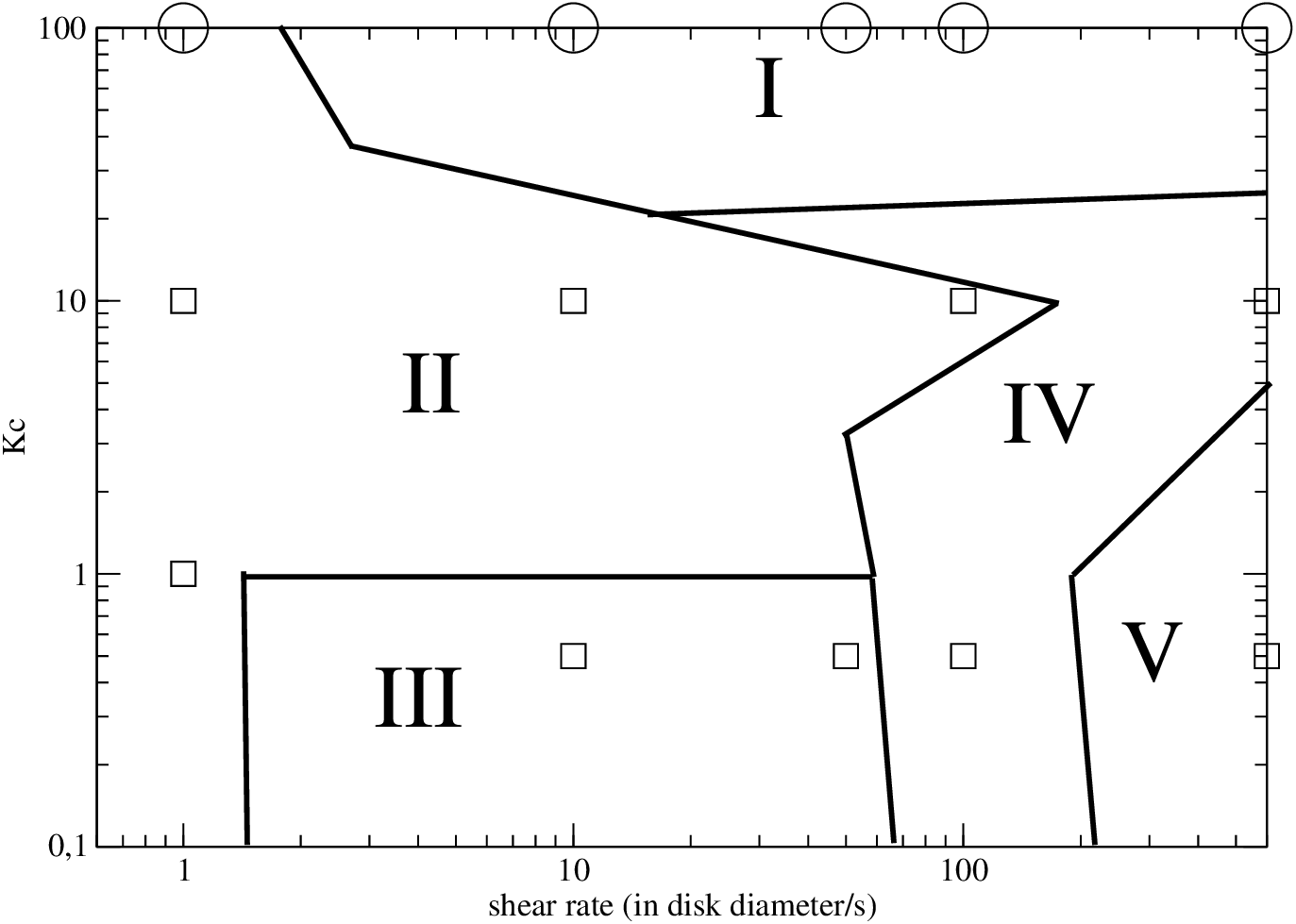}   
\caption{Diagram of the computed shear signal for a shear cell with blades and 
for non cohesive disks as a function of the elastic constant $K_c$ (in $g.s^{-2}$) of the cover and of the shear rate (in disk diameter per $s^{-1}$)}        
 \end{figure}

\begin{figure}  
 \includegraphics[width=14cm]{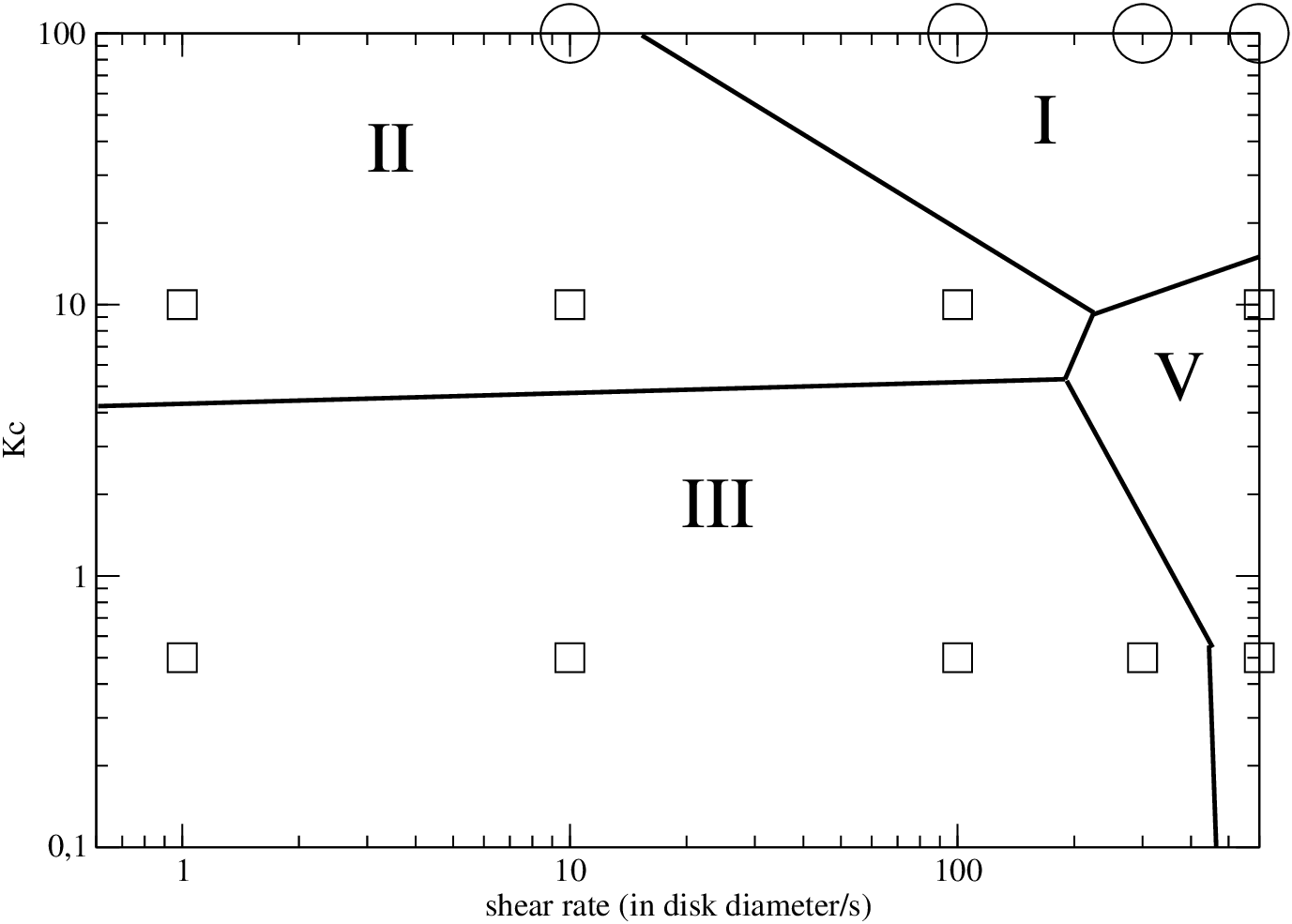} 
    \caption{Diagram of the computed shear signal for a shear cell with blades
 and for cohesive disks  as a function of the elastic constant $K_c$ ( $g.s^{-2}$) of the cover and of the shear rate (in disk diameter per $s^{-1}$)}       
      \end{figure}

\begin{figure}
\includegraphics[width=14cm]{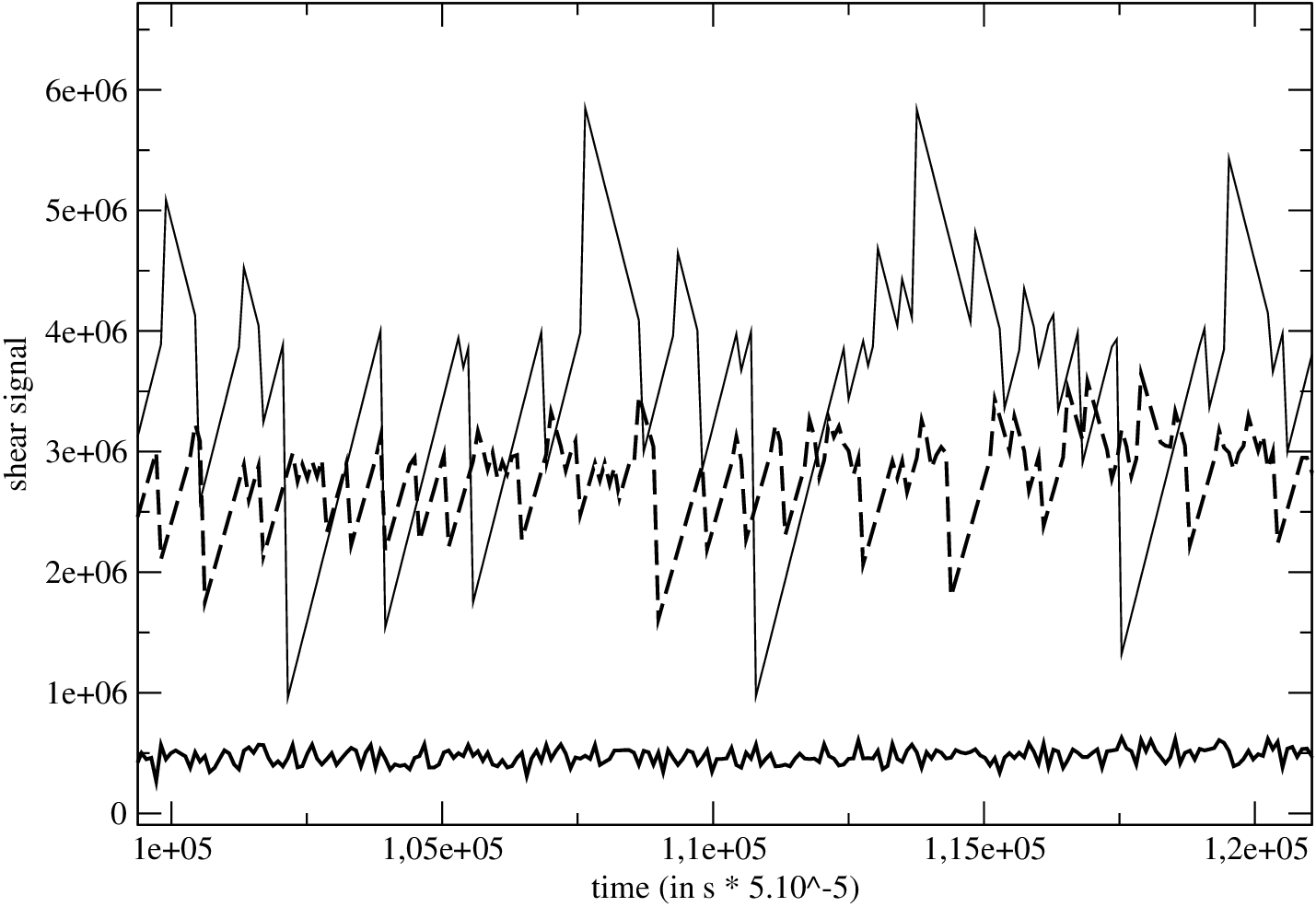}
\caption{Examples of computed shear signals as a function of time. 
Shear signal which is irregular in shape (continuous line), regular in shape
 but irregular in time (long dashed line), non linear in shape (thin continuous line). Experimental examples of similar behaviors may be seen in ref.\cite{oliver}.
}
\end{figure}

\end{document}